\pdfoutput=1

\documentclass{article}
\usepackage{fancyhdr}
\usepackage{spconf,amsmath,graphicx}

\usepackage{multirow}
\usepackage{amssymb}
\usepackage{hyperref}
\DeclareMathOperator*{\argmax}{argmax}
\DeclareMathOperator*{\argmin}{argmin}

\title{Zero-Shot Audio Classification with Factored Linear and Nonlinear Acoustic-Semantic Projections}
\name{Huang Xie$^{\star}$, Okko R\"as\"anen$^{\star \dagger}$, Tuomas Virtanen$^{\star}$}
\address{$^{\star}$ Unit of Computing Sciences, Tampere University, Finland \\
    $^{\dagger}$ Dept. Signal Processing and Acoustics, Aalto University, Finland}
\begin{document}
%\ninept
    \fancyhf{}

    \renewcommand{\headrulewidth}{0pt}

    \fancyfoot[c]{}

    \fancypagestyle{FirstPage}{
        \lfoot{\copyright 2021 IEEE. Personal use of this material is permitted. Permission from IEEE must be obtained for all other uses, in any current or future media, including reprinting/republishing this material for advertising or promotional purposes, creating new collective works, for resale or redistribution to servers or lists, or reuse of any copyrighted component of this work in other works.}
    }

    \setlength{\abovedisplayskip}{3pt}
    \setlength{\belowdisplayskip}{3pt}
    \maketitle
    \begin{abstract}
        In this paper, we study zero-shot learning in audio classification through factored linear and nonlinear acoustic-semantic projections between audio instances and sound classes.
        Zero-shot learning in audio classification refers to classification problems that aim at recognizing audio instances of sound classes, which have no available training data but only semantic side information.
        In this paper, we address zero-shot learning by employing factored linear and nonlinear acoustic-semantic projections.
        We develop factored linear projections by applying rank decomposition to a bilinear model, and use nonlinear activation functions, such as tanh, to model the non-linearity between acoustic embeddings and semantic embeddings.
        Compared with the prior bilinear model, experimental results show that the proposed projection methods are effective for improving classification performance of zero-shot learning in audio classification.
    \end{abstract}
    \begin{keywords}
        audio classification, zero-shot learning, acoustic-semantic projection
    \end{keywords}

    \section{Introduction}
    \label{sec:introduction}

    \thispagestyle{FirstPage}

    Supervised learning has been well-studied for tackling audio classification problems, such as acoustic scene classification~\cite{Barchiesi2015Acoustic} and environmental sound classification~\cite{Piczak2015Environmental, Piczak2015ESC}.
    To obtain classifiers with satisfactory performance, existing supervised learning techniques require large amounts of annotated training data from target sound classes, which is labor-intensive and costly to acquire.
    Moreover, with the increasing diversity of observed sound classes, it becomes even more challenging for humans to collect sufficient annotated training data for all possible sound classes.
    Recent work~\cite{Salamon2017Deep, Koluguri2020Meta, Shi2020FewShot, Chou2019Learning} in the audio recognition literature that deal with the lack of adequate training data mainly apply~\emph{data augmentation}~\cite{Brian2015ASoftware},~\emph{meta learning}~\cite{Finn2017Model} and~\emph{few-shot learning}~\cite{Wang2020Generalizing} methods.
    However, a certain amount of representative training data from target classes is still indispensable to make these methods work.
    Furthermore, to classify instances from novel classes, a supervised learning classifier would require retraining for the novel classes, which can be time-consuming and requires exhaustive parameter tuning.

    In this paper, we consider the extreme case of audio classification where target sound classes have no available training samples but only class side information (e.g., textual descriptions).
    This problem is generally referred to as the~\emph{zero-shot learning}~\cite{Palatucci2009Zero}, which has been increasingly studied in the context of image classification.
    In contrast to conventional supervised learning, zero-shot learning uses training data from only predefined classes (i.e., seen classes) to obtain classifiers that can be generalized to novel classes (i.e., unseen classes).

    There is only limited work that has been done for zero-shot learning in audio classification.
    Due to the lack of training data from unseen classes, class side information is used as a compensation for exploring the relationship between seen classes and unseen classes to make zero-shot learning possible.
    Prior work~\cite{Islam2019SoundSemantics, Xie2019Zero, Xie2020ZeroShotAudio} tackled zero-shot learning by leveraging semantic side information of sound classes, such as textual labels, with a two-phase learning process.
    First, intermediate-level representations were learned for audio instances and sound classes, respectively.
    Audio instances were embedded into a low-dimensional acoustic space with feature learning techniques, such as~\emph{VGGish}~\cite{Hershey2017CNN}.
    Sound classes were represented by word embeddings in a semantic space, which were extracted from their semantic side information with pre-trained language models, such as~\emph{Word2Vec}~\cite{Mikolov2013Efficient}.
    Then, an acoustic-semantic projection was learned to associate acoustic embeddings with semantic embeddings.
    \mbox{Islam et al.}~\cite{Islam2019SoundSemantics} employed a two-layer fully-connected neural network to model a nonlinear projection of acoustic embeddings onto semantic embeddings.
    In our previous work~\cite{Xie2019Zero, Xie2020ZeroShotAudio}, a bilinear model was used to learn a bidirectional linear projection between acoustic embeddings and semantic embeddings.
    Another prior work~\cite{Choi2019Zero} was conducted by integrating the acoustic embedding learning phase with the acoustic-semantic projection learning phase to optimize them holistically.
    Thus, a nonlinear acoustic-semantic projection was inherently built into their model.

    In this paper, we extend our previous work~\cite{Xie2019Zero, Xie2020ZeroShotAudio} by introducing matrix decomposition and nonlinear activation functions (e.g., tanh) to the bilinear model.
    We develop factored linear and nonlinear acoustic-semantic projections for zero-shot learning in audio classification.
    Experimental results show that the proposed projection methods are effective for improving classification performance of zero-shot learning in audio classification.

    The remainder of this paper is organized as follows.
    In Section~\ref{sec:zero-shot-learning}, we introduce the concept of zero-shot learning in audio classification.
    Then, we present the proposed factored linear and nonlinear projections in Section~\ref{sec:factored-linear-and-nonlinear-projections}.
    We describe the training algorithm in Section~\ref{sec:training-algorithm}, and discuss the experimental results in Section~\ref{sec:experiments}.
    Finally, we conclude this paper in Section~\ref{sec:conclusion}.

    \section{Zero-Shot Learning}
    \label{sec:zero-shot-learning}

    In this section, we introduce the concept of zero-shot learning via semantic side information in audio classification.
    We denote the audio sample space by $X$, the set of seen sound classes by $Y$, and the set of unseen sound classes by $Z$.
    Note that $Y$ and $Z$ are disjoint.
    Let $\theta(x) \in \mathbb{R}^{d_{a}}$ be the acoustic embedding of an audio instance $x \in X$ in an acoustic space, and $\phi(y) \in \mathbb{R}^{d_{s}}$, $\phi(z) \in \mathbb{R}^{d_{s}}$ be the semantic embeddings of sound classes $y \in Y$ and $z \in Z$ in a semantic space, respectively.
    We are given the training data $S_{tr}=\{(x_{n}, y_{n}) \in X \times Y | n=1,\dots,N\}$, where $x_{n}$ is an annotated audio sample belonging to a seen sound class $y_{n}$.
    In zero-shot learning, our goal is to learn an audio classifier $f:X \rightarrow Z$, which can predict the correct sound class for an audio instance $x \in X$, given its acoustic embedding $\theta(x)$ and the semantic embeddings $\phi(z)$ of every candidate sound class $z \in Z$.

    In prior work~\cite{Islam2019SoundSemantics, Xie2019Zero, Xie2020ZeroShotAudio, Choi2019Zero}, this is done via learning an acoustic-semantic projection $T:\mathbb{R}^{d_{a}} \rightarrow \mathbb{R}^{d_{s}}$ of acoustic embeddings onto semantic embeddings.
    Given an audio instance $x \in X$ belonging to a sound class $z_{x} \in Z$, it is generally assumed that the projected acoustic embedding $T(\theta(x))$ in the semantic space should be closer to the semantic embedding $\phi(z_{x})$ of its correct sound class $z_{x}$ rather than those of other sound classes.
    A similarity scoring function $F:\mathbb{R}^{d_{s}} \times \mathbb{R}^{d_{s}} \rightarrow \mathbb{R}$, as known as the~\emph{compatibility function}, is then defined to measure how similar/compatible a projected acoustic embedding and a semantic embedding are.
    To measure the similarity of two embedding vectors, the popular choices of $F$ could be Euclidean distance~\cite{Islam2019SoundSemantics}, cosine similarity~\cite{Choi2019Zero}, dot product~\cite{Xie2019Zero, Xie2020ZeroShotAudio}, etc.
    Therefore, the classifier $f:X \rightarrow Z$ is formulated as\footnote{Note that the $\argmin$ operation is used for other compatibility functions, such as Euclidean distance, etc.}
    \begin{equation}
        \label{eq:ZSL_classifier}
        z_{x} = f(x) = \argmax_{z \in Z} F(T(\theta(x)),\phi(z)).
    \end{equation}
    During the training stage, the projection $T$ is trained with audio samples $(x_{n}, y_{n}) \in S_{tr}$ such that
    \begin{equation}
        \label{eq:training_classifier}
        y_{n} = f(x_{n}) = \argmax_{y \in Y} F(T(\theta(x_{n})),\phi(y)).
    \end{equation}
    For prediction, an audio instance will be classified into the sound class, the semantic embedding of which is most compatible with its projected acoustic embedding.

    \section{Factored Linear and Nonlinear Projections}
    \label{sec:factored-linear-and-nonlinear-projections}

    In this section, we introduce our approach for developing factored linear and nonlinear acoustic-semantic projections for zero-shot learning in audio classification.
    First, we present the bilinear model used in our previous work~\cite{Xie2019Zero, Xie2020ZeroShotAudio}, on which we build factored linear and nonlinear acoustic-semantic projections.
    Then, we describe a factored linear projection obtained by applying matrix decomposition to the bilinear model.
    After that, we introduce the nonlinear projections derived from the factored linear projection by introducing nonlinear activation functions, such as tanh\@.

    \subsection{Bilinear Model}\label{subsec:bilinear-model}

    Inspired by prior work~\cite{Akata2016Label, Xian2016Latent} in computer vision, we employed a bilinear model in~\cite{Xie2019Zero, Xie2020ZeroShotAudio} to learn a bidirectional linear projection between acoustic embeddings and semantic embeddings.
    The audio classifier $f:X \rightarrow Z$ was then written in a bilinear form as
    \begin{equation}
        \label{eq:bilinear_model}
        f(x) = \argmax_{z \in Z} \theta(x)^{'} W \phi(z),
    \end{equation}
    where $W$ was the learned projection matrix.
    Considering a projection of acoustic embeddings onto semantic embeddings, $T$ could be formulated as
    \begin{equation}
        \label{eq:cross_domain_proj}
        T(\theta(x)) = W^{'} \theta(x).
    \end{equation}
    Thus, in~\eqref{eq:bilinear_model}, the dot product was inherently defined as the compatibility function $F$
    \begin{equation}
        \label{eq:dot_production}
        F(T(\theta(x)),\phi(z)) = T(\theta(x))^{'} \phi(z).
    \end{equation}

    \subsection{Factored Linear Projection}\label{subsec:factored-linear-projection}

    In the case where $d_{a}$ and $d_{s}$ are large, it would be valuable to decompose $W$ into a product of two low-rank matrices $U_{d_{a} \times r}$ and $V_{r \times d_{s}}$ to reduce the effective number of parameters in the bilinear model.
    Thus, we consider the rank decomposition $W=UV$ in~\eqref{eq:cross_domain_proj} to build a factored linear projection
    \begin{equation}
        \label{eq:linear_projection}
        T(\theta(x)) = (UV)^{'} \theta(x) = V^{'} U^{'} \theta(x).
    \end{equation}

    \subsection{Nonlinear Projection}\label{subsec:nonlinear-projection}

    Based on the linear projection~\eqref{eq:linear_projection}, we consider introducing nonlinear activation functions (e.g., tanh) into it to model the potential nonlinear relationship between acoustic embeddings and semantic embeddings.
    We apply a nonlinear activation function $t$ to $U^{'} \theta(x)$.
    Therefore, a nonlinear projection is formulated as
    \begin{equation}
        \label{eq:nonlinear_projection1}
        T(\theta(x)) = V^{'} t(U^{'} \theta(x)).
    \end{equation}

    We notice the fact that~\eqref{eq:nonlinear_projection1} also defines a two-layer fully-connected neural network\footnote{Note that the bias parameters of each layer are ignored from~\eqref{eq:nonlinear_projection1} and later formulas in this paper.} with input $\theta(x)$, layer weights $U^{'}$ and $V^{'}$.
    To describe deeper fully-connected neural networks, we can simply add more activation functions and matrices between $U^{'}$ and $V^{'}$.
    For example, a three-layer fully-connected neural network is formulated as
    \begin{equation}
        \label{eq:nonlinear_projection2}
        T(\theta(x)) = V^{'} t(Q~t(U^{'} \theta(x))).
    \end{equation}
    where the matrix $Q$ denotes the weight parameters of the second fully-connected layer.

    \section{Training Algorithm}
    \label{sec:training-algorithm}

    In this section, we introduce the algorithm for learning an acoustic-semantic projection $T$ with training data $S_{tr}$.
    Given an audio sample $(x_{n},y_{n}) \in S_{tr}$, we consider the task of sorting sound classes $y \in Y$ in descending order according to their compatibility values $F(T(\theta(x_{n})),\phi(y))$.
    Our objective is to optimize $T$ so that the correct class $y_{n}$ would be ranked at top of the sorted class list, i.e., having the maximal compatibility value for $x_{n}$.

    Let $r_{y}$ be the position index of a sound class $y$ in the sorted class list.
    We define $r_{y}=0$ when $y$ is sorted at the first position.
    By applying a ranking error function~\cite{Usunier2009Ranking}, we transform position index $r$ into loss $\beta(r)$:
    \begin{equation}
        \label{eq:ranking_error}
        \beta(r) = \sum_{i=1}^{r} \alpha_{i},
    \end{equation}
    with $\alpha_{1} \geq \alpha_{2} \geq \dots \geq 0$ and $\beta(0)=0$.
    Specifically, $\alpha_{i}$ denotes a penalty to a class losing a position from $i-1$ to $i$.
    In this paper, we follow previous work~\cite{Akata2016Label, Usunier2009Ranking} and choose $\alpha_{i}=1/i$.

    To learn an acoustic-semantic projection $T$ with audio samples $(x_{n},y_{n}) \in S_{tr}$, we minimize the weighted approximate-rank pairwise objective~\cite{Weston2011WSABIE}
    \begin{equation}
        \label{eq:weighted_ranking}
        \dfrac{1}{N} \sum_{n=1}^{N} \dfrac{\beta(r_{y_{n}})}{r_{y_{n}}} \sum_{y\in Y} \max \{0, l(x_{n},y_{n},y)\},
    \end{equation}
    with the convention $0/0=0$ when $y_{n}$ is top-ranked.
    In this paper, we define the hinge loss $l(x_{n},y_{n},y)$ as
    \begin{equation}
        \label{eq:hinge_loss}
        \begin{split}
            l(x_{n},y_{n},y) = \Delta(y_{n},y) &+ F(T(\theta(x_{n})),\phi(y)) \\ &- F(T(\theta(x_{n})),\phi(y_{n})),
        \end{split}
    \end{equation}
    where $\Delta(y_{n},y)=0$ if $y_{n}=y$ and 1 otherwise.
    The objective~\eqref{eq:weighted_ranking} is convex and can be optimized through stochastic gradient descent.
    To prevent over-fitting, we regularize~\eqref{eq:weighted_ranking} with L2 norms of parameter matrices in $T$.

    \begin{table}[!t]
        \centering
        \begin{tabular}{c||c||c}
            \hline
            Class Fold & Sound Class & Audio Sample \\
            \hline
            Fold0      & 104         & 23007        \\
            \hline
            Fold1      & 104         & 22889        \\
            \hline
            Fold2      & 104         & 22762        \\
            \hline
            Fold3      & 104         & 22739        \\
            \hline
            Fold4      & 105         & 21377        \\
            \hline
        \end{tabular}
        \caption{Class folds in the selected subset of AudioSet.}
        \label{tab:AudioSet_subset}
    \end{table}

    \section{Experiments}
    \label{sec:experiments}

    In this section, we evaluate the proposed method with AudioSet~\cite{Gemmeke2017AudioSet} and report the effectiveness of factored linear and nonlinear projections.

    \subsection{Dataset}\label{subsec:dataset}

    AudioSet~\cite{Gemmeke2017AudioSet} is a large unbalanced audio dataset, which contains over two million weakly labeled audio clips covering 527 sound classes.
    Most of these audio clips are multi-label.
    In this work, we focus on zero-shot learning in single-label classification problems.
    We follow the same experimental setup as in ~\cite{Xie2020ZeroShotAudio}.
    An audio subset containing 112,774 single-label 10-second audio clips and 521 sound classes is selected from AudioSet.
    We randomly split the selected subset into five disjoint class folds.
    The number of sound classes and audio samples in each class fold is shown in Table~\ref{tab:AudioSet_subset}.

    \subsection{Acoustic Embeddings}\label{subsec:acoustic-embeddings}

    A pre-trained VGGish~\cite{Hershey2017CNN} was used to generate acoustic embeddings from audio clips in~\cite{Xie2019Zero}.
    In zero-shot learning, it is generally assumed that unseen classes are unknown at training stage.
    Since a pre-trained VGGish may have already embedded knowledge about unseen sound classes, using it to generate acoustic embedding can lead to a biased evaluation of zero-shot learning in audio classification.
    Therefore, we train VGGish from scratch with audio data excluding unseen sound classes in this work.

    Following~\cite{Xie2019Zero, Xie2020ZeroShotAudio}, an 10-second audio clip is first split into ten one-second audio segments without overlapping.
    Then, a 128-dimensional embedding vector is generated for each audio segments with the trained VGGish.
    To obtain the clip-level acoustic embedding for an audio clip, we take the average of the 128-dimensional embedding vectors extracted from its one-second audio segments.

    \subsection{Class Semantic Embeddings}\label{subsec:class-semantic-embeddings}

    In AudioSet~\cite{Gemmeke2017AudioSet}, a sound class is described by one or several textual labels (i.e., words and phrases) and an additional short description (i.e., sentences).
    In this work, we consider only textual labels as class semantic side information.
    We adopt Word2Vec~\cite{Mikolov2013Efficient} as a word embedding model for generating semantic embeddings from these textual labels.
    For the sake of simplicity, we use a publicly available pre-trained Word2Vec\footnote{Word2Vec: https://code.google.com/archive/p/word2vec.}, which embeds roughly three million English words and phrases.
    It outputs a 300-dimensional semantic word vector for a single word or a phrase.
    To represent a sound class with semantic word vectors, we calculate the average of these word vectors extracted from its textual labels.

    \subsection{Experimental Setup}\label{subsec:experimental-setup}

    In the following experiments, we first train an VGGish for generating acoustic embeddings with class folds ``Fold0'' and ``Fold1''.
    Then, we conduct zero-shot learning in audio classification with ``Fold2'' for training, ``Fold3'' for parameter validation and ``Fold4'' for test, respectively.

    \textbf{VGGish Training}.
    Audio samples of class folds ``Fold0'' and ``Fold1'' are first randomly split into training/validation partitions with a class-specific proportion of 75/25.
    Then, we train an VGGish from scratch by feeding log mel spectrogram extracted from audio clips into it.
    After training, the trained VGGish achieves a classification accuracy (TOP-1) of 27.4\% on the validation partition.

    \textbf{Zero-Shot Learning}.
    We conduct zero-shot learning in audio classification with the proposed factored linear and nonlinear projections, respectively.
    We use the bilinear model as the baseline method.
    For the factored linear projection~\eqref{eq:linear_projection}, we experiment with low-rank decomposition and full-rank decomposition of $W$ to investigate the effect of the rank $r$ and the L2 norm regularization.
    For the nonlinear projection~\eqref{eq:nonlinear_projection1}, we experiment on three widely used activation functions, i.e., ReLU, sigmoid and tanh.
    We implement~\eqref{eq:nonlinear_projection1} by two-layer fully-connected neural networks, which are denoted by FC2$_{relu}$, FC2$_{sigmoid}$ and FC2$_{tanh}$, respectively.
    Similarly, we implement~\eqref{eq:nonlinear_projection2} by an three-layer fully-connected neural network with tanh, which is denoted by FC3$_{tanh}$.
    To prevent randomness, each projection method is evaluated twenty times with random initialization.
    The averages and standard deviations of their TOP-1 accuracies are reported in Table~\ref{tab:zsl_results}.

    \subsection{Result and Analysis}\label{subsec:result-and-analysis}

    As a baseline method, the bilinear model achieves an averaged TOP-1 of 5.7\% with a standard deviation of 1.1\%.
    For factored linear projections with either low-rank decomposition or full-rank decomposition of $W$, we obtain similar results (roughly 6.4 $\pm$ 0.6\%).
    We conclude that, with the L2 norm regularization, the rank $r$ has a limited influence on classification performance.
    Here, we report the result from a factored linear projection with the full-rank decomposition (i.e., \mbox{$r$=128}) of $W$, which has an averaged TOP-1 of 6.3\% with a standard deviation of 0.8\%.
    Unpaired t-test with $\alpha$=0.05 is used to measure the statistical significance among different methods.
    The results of the factored linear projection are significantly different from those of the bilinear model (\mbox{$t(38)$=2.09}, \mbox{$p$=0.04}).
    It shows that classification performance is improved by applying rank decomposition to $W$ with L2 norm regularization.
    For nonlinear projections, we set \mbox{$r$=128}.
    Classification performance is impaired with FC2$_{relu}$ (5.5 $\pm$ 0.9\%) while it is improved with FC2$_{sigmoid}$ (7.0 $\pm$ 0.5\%) and FC2$_{tanh}$ (7.2 $\pm$ 0.6\%).
    Particularly, the averaged TOP-1 of FC2$_{tanh}$ is significantly better than those of the bilinear model (\mbox{$t(38)$=5.60}, \mbox{$p$=$4.50\mathrm{e}{-6}$}) and the factored linear projection (\mbox{$t(38)$=3.88}, \mbox{$p$=$4.59\mathrm{e}{-4}$}).
    Compared with sigmoid and tanh, the ReLU function introduces non-linearity by simply dropping negative values from its inputs.
    In a two-layer fully-connected neural network, this can lead to a poor acoustic-semantic projection and results in an impaired performance.
    For FC2$_{sigmoid}$ and FC2$_{tanh}$, we think both of them capture non-linearity between acoustic embeddings and semantic embeddings, which is useful for improving classification performance.
    However, compared with FC2$_{tanh}$, classification performance is impaired with FC3$_{tanh}$ (6.0 $\pm$ 0.6\%).
    It seems that it would not be helpful for improving classification performance in zero-shot learning by simply introducing more nonlinear layers in a fully-connected neural network.

    \begin{table}[!t]
        \centering
        \begin{tabular}{c|l||c}
            \hline
            \multicolumn{2}{c||}{\bfseries Acoustic-Semantic } & \bfseries TOP-1 (\%) \\
            \multicolumn{2}{c||}{\bfseries Projection} & (avg $\pm$ std) \\
            \hline
            \multicolumn{2}{c||}{Bilinear (baseline)} & 5.7 $\pm$ 1.1 \\
            \hline
            \multicolumn{2}{c||}{Factored Linear} & 6.3 $\pm$ 0.8 \\
            \hline
            \multirow{4}{*}{Nonlinear} & FC2$_{relu}$    & 5.5 $\pm$ 0.9           \\
            \cline{2-3}
            & FC2$_{sigmoid}$ & 7.0 $\pm$ 0.5           \\
            \cline{2-3}
            & FC2$_{tanh}$    & \bfseries 7.2 $\pm$ 0.6 \\
            \cline{2-3}
            & FC3$_{tanh}$    & 6.0 $\pm$ 0.6           \\
            \hline
        \end{tabular}
        \caption{Zero-Shot learning in audio classification with different acoustic-semantic projections.}
        \label{tab:zsl_results}
    \end{table}

    \section{Conclusion}
    \label{sec:conclusion}

    In this paper, we present an approach for zero-shot learning in audio classification with factored linear and nonlinear acoustic-semantic projections.
    We develop factored linear and nonlinear projections by applying rank decomposition and nonlinear activation functions to a bilinear model.
    We evaluate our proposed approach with a large unbalanced audio dataset.
    The experimental results show that both factored linear and nonlinear projections are effective for zero-shot learning in audio classification.
    With a factored linear projection, it achieves an averaged TOP-1 accuracy of 6.3\%, which is better than the prior bilinear model (5.7\%).
    By introducing nonlinear activation functions into it, classification performance can be further improved.
    Classification performance achieves an averaged TOP-1 accuracy of 7.2\% by using a nonlinear projection with the tanh activation function.

    \section{Acknowledgement}
    \label{sec:acknowledgement}

    The research leading to these results has received funding from the European Research Council under the European Unions H2020 Framework Programme through ERC Grant Agreement 637422 EVERYSOUND. OR was funded by Academy of Finland grant no. 314602.

% References should be produced using the bibtex program from suitable
% BiBTeX files (here: strings, refs, manuals). The IEEEbib.bst bibliography
% style file from IEEE produces unsorted bibliography list.
% -------------------------------------------------------------------------
    \bibliographystyle{IEEEbib}
    \bibliography{strings,refs}

\end{document}